\RequirePackage{fix-cm}
\documentclass[smallextended]{svjour3}       
\smartqed  

\usepackage{graphicx}
\usepackage{epstopdf}
\usepackage{Algorithm}
\usepackage[noend]{Algorithmic}
\usepackage{subfigure}
\usepackage{color}

\newcommand{\LINEIF}[2]{%
    \STATE\algorithmicif\ {#1}\ \algorithmicdo\ {#2}%
}

\newcommand\NoThen{\renewcommand\algorithmicthen{}}

\begin{document}

\title{Asynchronous Liquids: Regional Time Stepping for Faster SPH and PCISPH}


\author{Prashant Goswami*         \and
        Christopher Batty 
}


\institute{P. Goswami \textit{(*corresponding author)} \at
              Blekinge Institute of Technology, Sweden \\
              Tel.: +0 (46) 455 - 38 58 26 \\
              Fax: +0 (46) 455 - 38 50 57\\
              \email{prashant.goswami@bth.se}           
           \and
              C. Batty \at
              University of Waterloo, Canada \\
              Tel.: +1 (519) 888-4567 \\
              Fax:. +1 (519) 885-1208 \\
              \email{christopher.batty@uwaterloo.ca}
}

\date{Received: date / Accepted: date}

\maketitle

\begin{abstract}
This paper presents novel and efficient strategies to spatially adapt the amount of computational effort applied based on the local dynamics of a free surface flow,
for both classic weakly compressible SPH (WCSPH) and predictive-corrective incompressible SPH (PCISPH). Using a convenient and readily parallelizable block-based approach,
different regions of the fluid are assigned differing time steps and solved at different rates to minimize computational cost. Our approach for WCSPH scheme extends an
asynchronous SPH technique from compressible flow of astrophysical phenomena to the incompressible free surface setting, and further accelerates it by entirely decoupling
the time steps of widely spaced particles. Similarly, our approach to PCISPH adjusts the the number of iterations of density correction applied to different regions, and
asynchronously updates the neighborhood regions used to perform these corrections; this sharply reduces the computational cost of slowly deforming regions while preserving
the standard density invariant.We demonstrate our approaches on a number of highly dynamic scenarios, demonstrating that they can typically double the speed of a simulation
compared to standard methods while achieving visually consistent results.


\end{abstract}


\section{Introduction}

The \textit{Smoothed Particle Hydrodynamics (SPH)} method is a powerful and widely used approach to liquid animation \cite{Muller:2003:PFS:846276.846298,Monaghan:2005,Becker:2007:WCS:1272690.1272719};
among other benefits, it produces detailed splashing and droplet effects, supports seamless topological changes and preservation of liquid mass, and handles complex boundaries in a
straightforward manner. However, capturing a sufficiently wide range of spatial scales in order to generate visually compelling results often requires large particle counts, and
correspondingly long simulation times.

To date, several acceleration strategies have been proposed to tackle this challenge, including GPU or multi-core CPU methods that exploit parallelism (e.g., \cite{Goswami:2010:ISS:1921427.1921437,IhmsenABT11:journals,vriphys.20151331})
and spatially adaptive methods that coarsen the particle's spatial resolution away from the surface (e.g., \cite{Adams:2007:ASP:1276377.1276437,Solenthaler:2011:TPS:2010324.1964976}).
We propose a new and complementary approach.

Across all SPH methods, the choice of time step remains a crucial factor in determining the overall computational cost. All else being equal, the smaller the time step, the more iterations
that must be taken to simulate a given span of time, and hence the longer the total time spent running the simulation. The standard time stepping strategy is to use a single global time step
which is either held constant throughout the simulation, or varied as a function of the most rapidly deforming region of the flow to ensure stability and accuracy \cite{IhmsenAGT10}. However,
 many practical fluid flows involve both slow movement and comparatively rapid movement, due to external forces, inflow/outflow boundaries, collisions with objects, and so forth. A global
 time step is often \emph{much too conservative} in slow moving regions, leading to a great deal of wasted computational effort where a large time step would suffice.

Problems of this nature suggest the use of \emph{asynchronous time integration}: different regions of a simulation should be computed at different rates in order to maximize efficiency
while satisfying accuracy and stability restrictions. Variations on this idea have been applied to animation problems in rigid bodies, cloth, deformable bodies, and collision processing
\cite{Mirtich:2000:TRB:344779.344866,Thomaszewski2008,Harmon:2009:ACM:1576246.1531393}, and it has a long history in mechanics (e.g. \cite{Belytschko1981}). This general strategy has also
been developed for certain SPH simulations in astrophysics \cite{Owen1998,Serna:2003} and weakly compressible SPH (WCSPH) \cite{Becker:2007:WCS:1272690.1272719} in
\cite{egsh.20141011,egsh.20151010}. However, to our knowledge this concept has not been extended to animating free-surface flow of incompressible SPH methods like (PCISPH)
scheme \cite{Solenthaler:2009:PIS:1531326.1531346}.

In this paper, we introduce \emph{regional time stepping} (RTS) approaches for both the WCSPH and PCISPH methods, in which computational effort is expended on different fluid regions
in proportion to the speed of their local dynamics. In our numerical experiments, we were able to reduce simulation times by approximately a factor of two compared to global adaptive
time stepping on realistic, highly dynamic scenes in which the entire connected body of fluid is in motion. Our algorithm relies on an efficient block-based technique to determine the
different regions and support convenient parallelism. After reviewing related work, we will outline how to choose the regions and their corresponding time steps, and then describe how
we effectively incorporate this central idea into each of the WCSPH and PCISPH schemes.

This paper is an extension of our earlier published short paper on RTS for WCSPH \cite{egsh.20141011}. Here we extend our work to incorporate RTS to predictive-corrective incompressible SPH and
also provide more detailed exposition of our method.

\section{Related Work}

The smoothed particle hydrodynamics method, or SPH, was first applied to liquid animation by M\"{u}ller et al.\ \cite{Muller:2003:PFS:846276.846298}, although Desbrun and
Cani \cite{DC1996,DC99b} had earlier applied it to animating highly deformable bodies. Further background on the classic weakly compressible SPH scheme can be found in a
review by Monaghan \cite{Monaghan:2005} and a paper by Becker and Teschner \cite{Becker:2007:WCS:1272690.1272719}. More recently, the predictive-corrective incompressible
variant of SPH (PCISPH) introduced by Solenthaler and Pajarola \cite{Solenthaler:2009:PIS:1531326.1531346} has been widely adopted because it allows for significantly larger time
steps while maintaining incompressibility.

\begin{figure}[t]
   \centering
     \includegraphics[width=.49\linewidth]{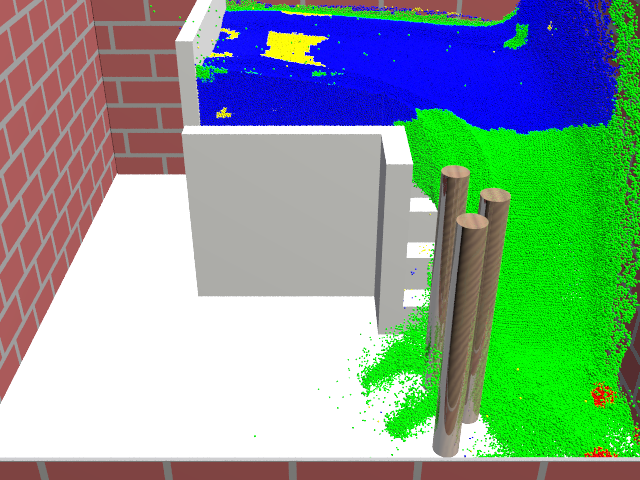}
     \includegraphics[width=.49\linewidth]{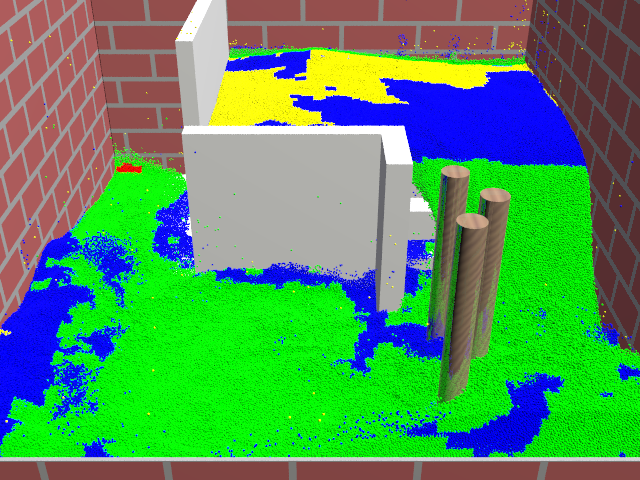}
     \includegraphics[width=.49\linewidth]{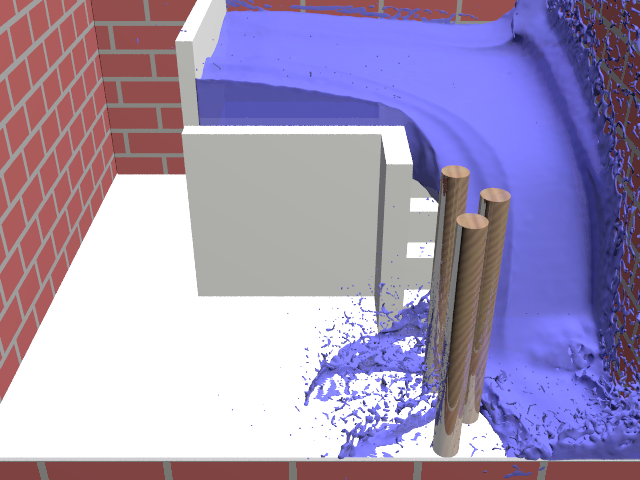}
     \includegraphics[width=.49\linewidth]{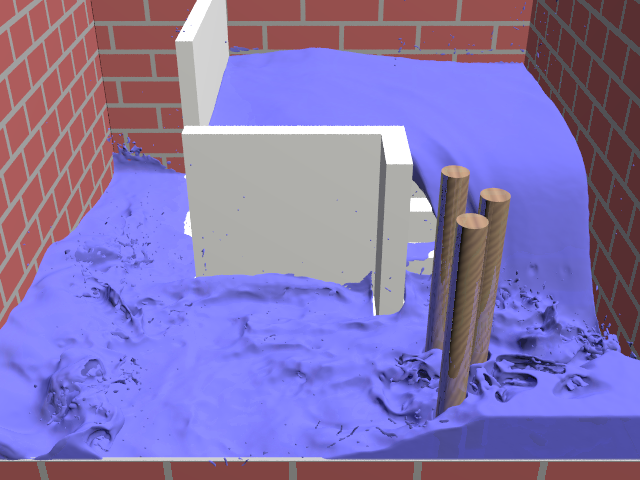}
     \includegraphics[width=.49\linewidth]{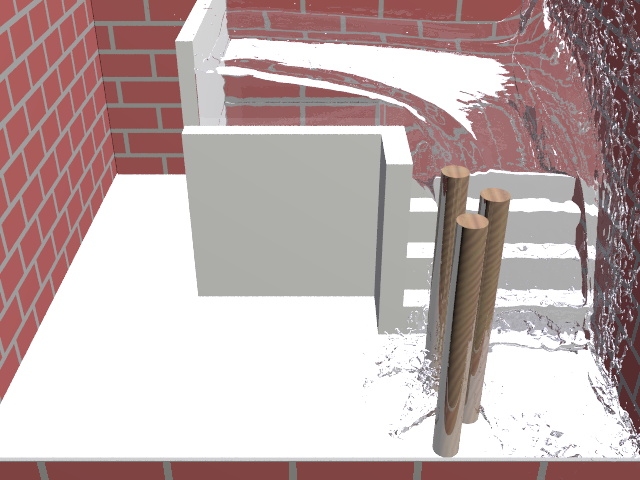}
     \includegraphics[width=.49\linewidth]{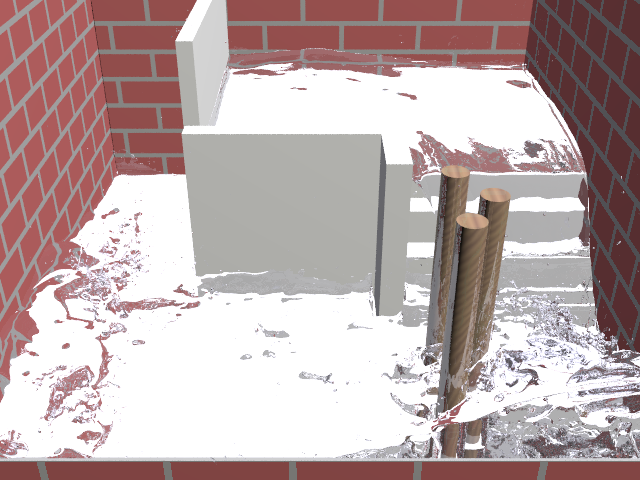}
  \caption{Water gushing down stairs and along a corridor, simulated using our regional time stepping PCISPH method with 1.6M particles. We achieve a factor of 3.6 speed-up
  over a constant time step approach method and a factor of 1.7 speed-up over global adaptive time stepping [Ihmsen et al. 2010]. Top: Particles colored by time step region.
  Middle: Opaque surface geometry. Bottom: Refractive water rendering.}
   \label{fig:pcisph_corridor}
\end{figure}

The Navier-Stokes equations governing fluid flow naturally yield behavior spanning a wide range of spatial and temporal scales; depending on the application of interest,
certain of these features are more relevant than others. For example, in liquid animation the surface motion and details often take priority, and this is captured in spatially
 adaptive approaches that coarsen the particle scale further from the surface to reduce the total number of particles \cite{Adams:2007:ASP:1276377.1276437,Zhang:2008:ASR:2386410.2386433,Solenthaler:2011:TPS:2010324.1964976}.
 On the other hand, both GPU-based methods \cite{Harada:2007:CGI,Zhang:2008:ASR:2386410.2386433,Goswami:2010:ISS:1921427.1921437} and multi-core CPU-based
 methods \cite{IhmsenABT11:journals} have also been proposed to accelerate SPH simulations, without necessarily relying on adaptivity. The application of asynchronous
 time stepping is largely orthogonal to many of these approaches, and therefore complementary; we demonstrate our new method within a parallel CPU-based SPH code.

As noted earlier, the time step is a crucial factor in the computational cost of SPH. The most common strategy to incorporate temporal adaptivity is to modify the global
time step over the course of the simulation based on the maximum velocities and forces of the entire fluid body at a given time. This allows the simulation to proceed more
rapidly during calm motions, but to take much smaller time steps when necessary to resolve very rapid motion. For example, this strategy was recently adapted to the PCISPH
method by Ihmsen et al.\cite{IhmsenAGT10}. Raveendran et al.\cite{Raveendran:2011:HSP:2019406.2019411} proposed a rather different multi-resolution strategy to
allow large time steps: the SPH method is augmented with a broad-scale Eulerian projection method to provide a good initial guess at the fluid pressure. In contrast, our
approaches are purely Lagrangian.

We are aware of no methods for liquid animation that exploit the possibility of varying the time step itself spatially. One partial exception is the method of Goswami and
Pajarola \cite{GoswamiP11} in which very slow moving particles are entirely frozen to save computational cost. Although this accelerates the simulation, its applicability
is fairly limited and it can introduce objectionable dissipation effects in the fluid motion if applied aggressively.

SPH methods in astrophysics applications have employed asynchronous time integration strategies \cite{Hernquist1989,Owen1998,Serna:2003,GRADSPH} to deal with large
variations in time scales and stiffnesses. This setting differs from ours in that the target medium is typically compressible and doesn't involve a free-surface. Our asynchronous
approach for weakly compressible SPH builds on that of Serna et al.\cite{Serna:2003}, augmenting it with a block-based approach that lets us smooth temporal variations
between regions and skip a larger amount of computation in less active regions. Furthermore, we develop a novel regional time stepping method for PCISPH that extends many of the
advantages of asynchrony to this setting as well.

Lastly, we note that while projection-based Eulerian methods for incompressible flow are inherently synchronous to some degree, Patel et al.\cite{Patel2005} explored
using distinct time steps for disjoint liquid bodies of the same simulation to gain some of the benefits of asynchrony.

\section{Block-based Computation}
\label{section:blocks}

Our algorithm relies on a block-based architecture. If $s$ is the initial particle spacing, we divide the simulation domain into a virtual grid, with each block having
support radius $r$, such that $r \simeq 2s$. Thus each particle is contained by exactly one of the blocks in the simulation domain.

Such an arrangement has several benefits. For example, neighbors of all particles in a block can be computed efficiently by examining neighboring blocks. Each block can
also be treated as a parallelization unit for computing the physics of particles within it, as in the work of Goswami et al.\cite{Goswami:2010:ISS:1921427.1921437}.

However, the most important advantage of the block-based arrangement in our case is parallel region determination. The time steps for a given region are computed over these
virtual blocks instead of at the particle level, under the reasonable assumption that liquid in a local area tend to be deforming at comparable rates. This method can then be
efficiently parallelized by launching a thread per filled block instead of per particle. Particles falling within that block report their velocity and force up to the parent block,
thereby avoiding any race or collision conditions.

\subsection{Time Step Selection}

\begin{figure}[ht]
  \centering
  \includegraphics[width=1.0\linewidth]{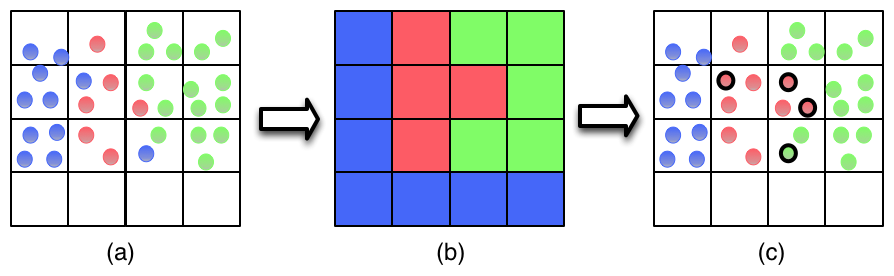}
  \caption{Block-based time step computation for particles, red ($\Delta t_b$), green ($2\Delta t_b$) and blue ($3\Delta t_b$) where $\Delta t_b$ is the base (smallest) time step.
  (a) Particles colored by the individual time steps they would ordinarily possess. (b) Particles pass their velocity and force values to their parent blocks. Each block is assigned
  the minimum required time step based on its particles. (c) The block propagates its computed time step back to the particles. (Particles whose time step has been altered are outlined
  in black.) }
  \label{fig:blocks}
\end{figure}

Our simple block-based time step computation is illustrated in Figure \ref{fig:blocks}, and comprises three steps:
\begin{enumerate}
\item All particles compute their velocity and total force.
\item Particles propagate their attributes to their parent (i.e., containing) block. A minimum time step is computed for the block based on the maximum force and velocity from its particles.
\item Each block's time step is propagated back to its particles.
\end{enumerate}

This approach is used both for WCSPH and PCISPH.

In what follows, $\Re_n$ denotes a \emph{region} or set of blocks assigned to a given time step $\Delta t_n = n \Delta t_b$ where $\Delta t_b$ is the base time step, and $n$ is a positive
integer. The corresponding particle set is denoted by $S_n$.

A block is assigned to a region corresponding to the largest time step for which it satisfies a set of three criteria, according to its particles' maximum velocity and force. The first
two criteria are:
\begin{equation}
\Delta t_n  \le \frac{\lambda_v r} {c_s}
\label{eq:timeSound}
\end{equation}

\begin{equation}
\Delta t_n \le \lambda_f \sqrt{\frac{rm}{F_{max}}}
\label{eq:force}
\end{equation}

These are standard time step conditions from the SPH literature (e.g., \cite{DC99b,Becker:2007:WCS:1272690.1272719}); the first is a CFL condition, while the second accounts for sudden
accelerations over a time step. In these equations, $c_s$ is the speed of sound in the medium, $m$ is the particle mass, $F_{max}$ is the maximum force magnitude of particles in the block,
and $V_{max}$ is the maximum velocity magnitude of particles in the block. We set the remaining coefficients $\lambda$ to $\lambda_v \le 0.4$ and $\lambda_f \le 0.25$.

We introduce a third criterion to partition particles into groups depending on their velocities:
\begin{equation}
\frac{\Delta t_n V_{max}}{r} \le \alpha \beta_{n}
\label{eq:time}
\end{equation}

In essence, Equation \ref{eq:time} assigns to each particle a time step based on the fraction of its support radius that it would cover in a step moving at its current velocity.
The threshold cutoffs $\beta_n$ determine how the time steps are partitioned.

For SPH all three criteria are applied, and we set $\alpha=0.4$. For PCISPH, larger time steps can safely be taken than the classic CFL condition would dictate, so equation
\ref{eq:timeSound} is omitted when assigning blocks to regions, and we set $\alpha=1$.

In our implementation, the value of $\beta$ was set as follows:

$\beta_1$ = $\infty$, \hspace{1mm}
$\beta_n$ = $0.4  {(0.2)}^{(n-2)}$ for $n \ge 2$

By assigning an arbitrarily large value to $\beta_1$ we ensure that $\Re_1$ is assigned the smallest time step.
The choice of $\beta_2$ comes directly from the CFL condition, and higher coefficients are obtained by scaling down the previous value.

\section{Regional Time Stepping with WCSPH}

\subsection{Individual Time Stepping for SPH}

Serna et al.\cite{Serna:2003} introduced an asynchronous predictor-corrector time integration strategy for their DEVA astrophysical SPH code, later also used by the
GRADSPH code \cite{GRADSPH}. We begin by briefly reviewing this method, and refer readers to Serna's work for an expanded exposition.

Given a set of particles assigned different time steps, consider advancing through the union of all the resulting time steps.
Beginning from a current time $t_n$, with positions $x^n_i$, velocities $v^n_i$, and accelerations $a^n_i$ for each particle $i$, the following predictor step of length
$\Delta t = t_{n+1}-t_n$ is taken by \emph{all} particles to estimate new velocities and positions at time $t_{n+1}$:
\begin{eqnarray}
\tilde{x}^{n+1}_{i} &=& x^n_i + v^n_i \Delta t + \frac{a^n_i(\Delta t)^2}{2} \\
\tilde{v}^{n+1}_{i} &=& v^n_i +  a^n_i \Delta t
\end{eqnarray}
Among the set of all particles, the time $t_{n+1}$ will be the conclusion of a ``true" time step for some, called \emph{active particles}; for the remainder this step is taken
only to provide intermediate information to nearby particles.
Next, only the active particles have their neighborhoods and accelerations re-evaluated at $t_{n+1}$, and their positions and velocities are corrected:
\begin{eqnarray}
x^{n+1}_{i} &=& \tilde{x}^{n+1}_{i} + \frac{(a^{n+1}_i - a^n_i)\delta t^2}{6} \\
v^{n+1}_{i} &=& \tilde{v}^{n+1}_{i} + \frac{(a^{n+1}_i - a^n_i)\delta t}{2}
\end{eqnarray}
Crucially, $\delta t$ refers to the length of time between $t_{n+1}$ and the last time that each specific particle's acceleration was evaluated (i.e., the length of its true time step).
All other particles maintain their previous acceleration value.

The net effect is that particles taking large time steps assume constant acceleration over their true step, and the intermediate predictor steps approximate the necessary ``substep"
information required by nearby particles that may be taking smaller (or offset) time steps. Because acceleration is assumed constant (and position and velocity treated accordingly),
the number of substeps taken does not change the final end-of-step positions or velocities for the particles being substepped, compared to taking a single large step; the substepping
is merely an interpolation process.

\begin{figure}[t]
   \centering
     \includegraphics[width=.6\linewidth]{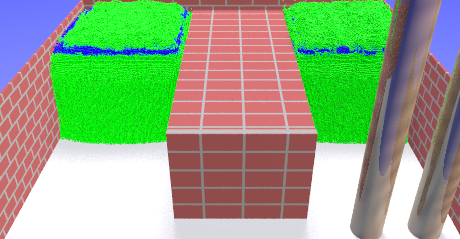}
     \includegraphics[width=.6\linewidth]{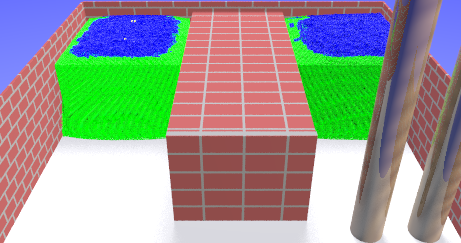}
   \caption{The initial moments of the double dam break example using RTS WCSPH. Top: Without applying the second order velocity and position corrections of Serna et al., the particles
   quickly push apart and the smooth surface is disrupted. Bottom: With the corrections applied we achieve the expected behavior.}
   \label{fig:sph_correctionSerna}
\end{figure}

The correction applied at the end of a particle's true time step maintains second order accuracy in position and velocity. Furthermore, without this correction na\"{i}ve asynchronous
simulations exhibit visual artifacts. As evident from the noisier surface and altered color distribution in the dual dam breaking scenario in Figure \ref{fig:sph_correctionSerna}, the
particles' natural motion and stability is disrupted, erroneously leading to smaller time steps.

\subsection{Incorporating Regional Time Steps}
Although this approach saves on expensive evaluations of forces and accelerations, it still requires substepping of \emph{all} particles in the simulation at the smallest global time step.
We make the further observation that if all the particles within a given particle's neighborhood require only the same or larger time step, then no interpolated substeps need to be taken
and the final result will be the same. Therefore in regions of our domain assigned large time steps, we can safely integrate all the contained particles at that timestep without the need
to perform any substepping whatsoever. This allows the simulation to remain synchronized overall, while correctly integrating different regions at appropriate rates and avoiding unnecessary
computation.

The basic outline of our approach is presented in Algorithm \ref{alg:SPH_RTS}.
The first step assigns a time step to each block within the simulation domain. That is, we choose $\Re_n$ and $S_n$ and update the global block-based neighborhood grid.

To ensure that the time step varies gradually across the physical domain, which aids in simulating quite stiff incompressible flows, we locate the boundary between regions with different
time steps, and determine the set of blocks $\Re_{min}$ on the side with the larger time step. This region is then assigned the smaller time step of its neighboring region, which is done
efficiently at the block level by checking each block's neighbors.

In our algorithm, the particles maintain a few additional variables. $validity$ is the number of the smallest time steps for which its most recently computed attributes are assumed
valid (i.e., how many substeps before its true time step ends). $compute$ is a Boolean flag that indicates whether the particle is currently active (i.e., requires re-computation of its
acceleration, and end-of-step correction of its position and velocity) which occurs when the $validity$ $\le$ 0. If the particle is active, its neighborhood set is determined and its
local density and forces are computed. Otherwise, it skips these steps. At the end of each loop, the position and velocity of each particle is updated, and active particles have their
velocities and positions corrected (lines 25-30), per Serna's scheme \cite{Serna:2003}.

We make some additional observations. First, while the computation of time steps is determined per block, it is updated on the individual particles which also track their own validity.
Blocks do not have validity or history, and therefore all computations over blocks are valid only for a frame.
Second, the algorithm is pre-emptive. That is, a particle can change its time step even before its validity expires (line 14 of Algorithm \ref{alg:SPH_RTS}).
This allows the method to maintain stability in the face of sudden accelerations, as often occurs in collisions with boundaries.

\begin{algorithm}
\begin{algorithmic}[1]
\FORALL {particles $i$}
    \STATE {set $validity_i$ = $0$} \newline
\ENDFOR

\WHILE{(animating)}
    \STATE{update block neighborhood grid} \newline

    \STATE{\textcolor{blue}{/*--------- Region determination($\Re_i$) ---------*/}}
    \FORALL{$i$ $\in$ $S$}
        \STATE {update parent block maximum with $v_i$ and $F_i^{total}$} \newline
     \ENDFOR

    \FORALL{blocks $b$}
        \STATE {compute new region membership per section \ref{section:blocks}} \newline
     \ENDFOR

     \FORALL{$i$ $\in$ $S$}
        \STATE {decrement $validity_i$}
        \NoThen
        \IF {($validity_i$ $\leq$ $0$) $\|$
             (\textit{parent block has a different time step})}
            \STATE {set $compute_i$ $=$ \textbf{true}}
            \STATE {update $validity_i$, $timestep_i$}
        \ELSE
            \STATE {$compute_i$ $=$ \textbf{false}} \newline
        \ENDIF
     \ENDFOR

    \STATE{\textcolor{blue}{/*--------- Physics computation ---------*/}}
        \FORALL{$i$ $\in$ $S$}
            \LINEIF{$compute_i$}{find neighborhoods $N_{i}$} \newline
        \ENDFOR

        \FORALL{$i$ $\in$ $S$}
            \LINEIF{$compute_i$}{update $\rho_i$, $p_i$} \newline
        \ENDFOR

        \FORALL{$i$ $\in$ $S$}
            \LINEIF{$compute_i$}{update $\textbf{F}^{p, v, g}$} \newline
        \ENDFOR

     \FORALL{$i$ $\in$ $S$}
        \STATE {predict new $v_i$}
        \STATE {predict new $x_i$}
        \IF{$compute_i$}
            \STATE {apply correction to $v_i$}
            \STATE {apply correction to $x_i$}
        \ENDIF
     \ENDFOR

\ENDWHILE
\end{algorithmic}
\caption{Regional Time Stepping for WCSPH}
\label{alg:SPH_RTS}
\end{algorithm}

\begin{figure}[t]
   \centering
     \includegraphics[width=.49\linewidth]{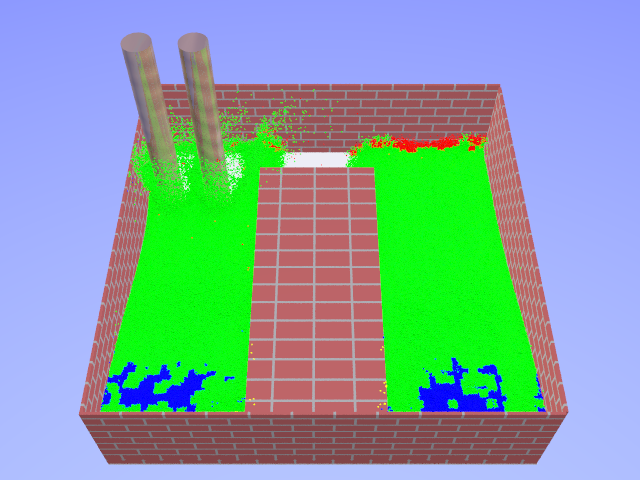}
     \includegraphics[width=.49\linewidth]{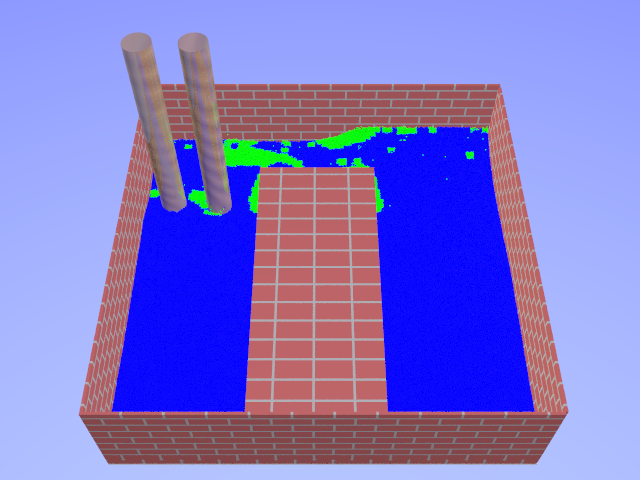}
     \includegraphics[width=.49\linewidth]{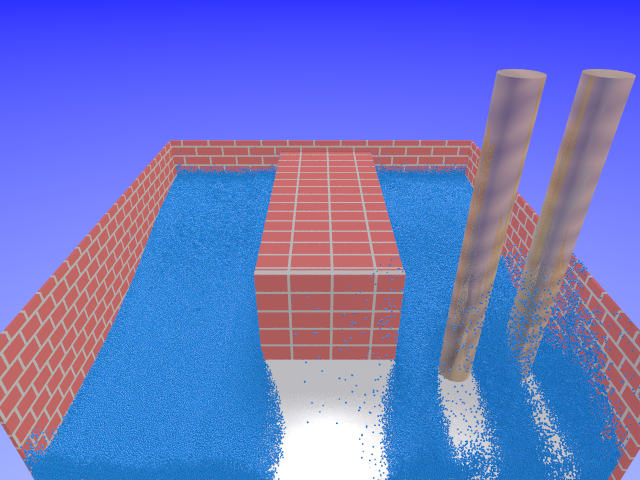}
     \includegraphics[width=.49\linewidth]{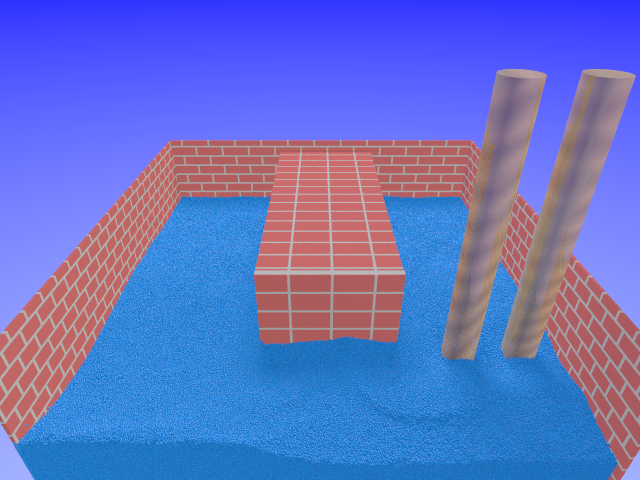}
  \caption{A 1.5M particle double dam break simulated using our regional time stepping algorithm for WCSPH. Top: Particles colored by time step region. Bottom: Particles for corresponding
  frames from a different viewpoint and colored uniformly.}
   \label{fig:sph_doubleDB}
\end{figure}

\section{Regional Time Stepping with PCISPH}
\subsection{Motivation}
The second major contribution of our work is to develop a regional time stepping method for predictive-corrective incompressible SPH \cite{Solenthaler:2009:PIS:1531326.1531346}. As the name
suggests, PCISPH enforces incompressibility through a predictor-corrector approach, which iteratively refines pressure forces to correct any deviations in particle density.
This allows for time steps about an order of magnitude larger than weakly compressible SPH, while recovering near-identical behavior. To accelerate this method, Ihmsen et al.\cite{IhmsenAGT10}
proposed an adaptive time stepping PCISPH scheme that adjusts the global time step depending on the simulation state. This does indeed improve the speed of PCISPH, however its overall
efficiency is limited by fast-moving regions, which can arise frequently during collisions with boundaries.

Similar to our method for WCSPH, our essential observation is that slow moving regions should require less computational effort to simulate a given amount of time. Concretely, for PCISPH,
this is because the particle density changes more slowly in these regions, and therefore these density variations ought to require fewer corrective iterations to resolve. (In their adaptive
time stepping work, Ihmsen et al.\ noted the converse: fast motion and large impacts can require many more density correction iterations.)

The second observation we build on is that more localized density corrections can be highly effective. Raveendran et al.\cite{Raveendran:2011:HSP:2019406.2019411} noticed this, and exploited
it by applying a post-process that spends extra iterations correcting only those particles with large remaining density errors after their core algorithm concludes. We instead make this
observation a fundamental feature of our algorithm, locally applying a different number of density correction iterations based on the dynamics of different regions of the flow.

\subsection{The Algorithm}
At a high level, our algorithm works as follows: we pick a large time step $t_n$, called the major time step, and divide it into $n$ equal subintervals, or minor steps, so that
$t_n = n \Delta t_b$. (We used $n=4$.) At the beginning of a major step, we assign blocks of particles to different regions based on their dynamics as in section \ref{section:blocks}.
On each minor step, all regions perform at least one iteration of density correction, and then participate in additional iterations depending on their region membership.

Note that our algorithm is therefore not truly asynchronous in the manner of our RTS SPH approach; all particles are advanced in synchronization. However, the computational expense of
slow moving regions is dramatically reduced, by lowering the number of correction iterations applied. On the other hand, we do update the particle neighborhoods asynchronously in proportion
to how fast they are likely to change, since neighborhood searches are a major expense in SPH algorithms. As in WCSPH, we maintain a $validity$ variable for each particle that tracks how
many (minor) time steps a particle's data is considered to be valid for, based on its region membership. Primarily, this means that we update the particle neighborhood only when its validity
expires, and otherwise reuse its most recently computed neighborhood.

Pseudocode for our approach is given in Algorithm \ref{alg:PCISPH_RTS}, and we describe its various elements below.

 \begin{figure}
   \centering
   \includegraphics[width=0.7\linewidth]{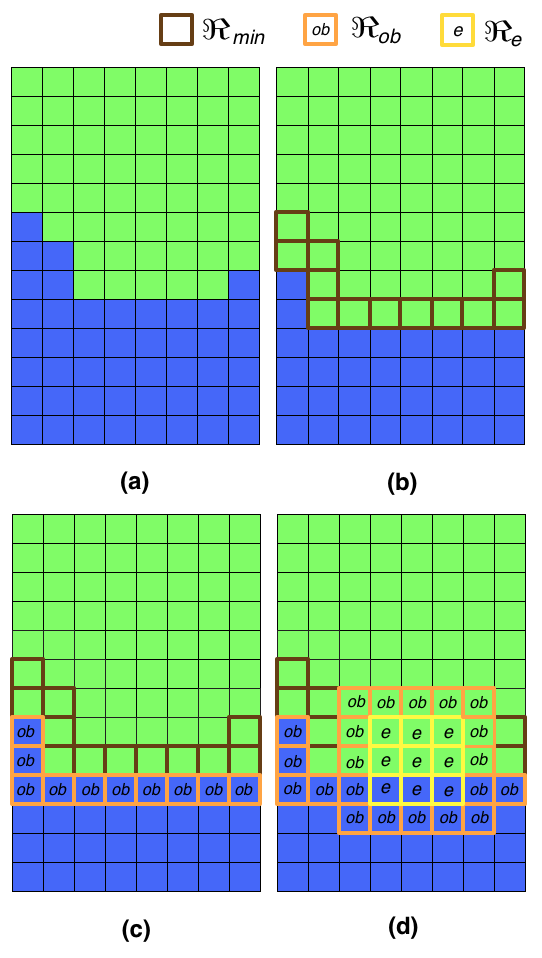}
   \caption{Block-based region determination. (a) Blocks are assigned time steps. (b) At borders between time step regions, we assign blocks with larger time steps (blue, $3t$) the time step
   of their neighbour with a smaller time step (green, $2t$) to smooth out region transitions. This region is called $\Re_{min}$, shown with brown borders. (c) For RTS PCISPH, we must also
   identify the set of blocks $\Re_{ob}$ which border region changes on the side with the larger time step; this region is observed for density variations of particles that were previously
   resolved. (d) We also identify the region $\Re_{e}$ containing particles with remaining density errors regardless of their region, and augment $\Re_{ob}$ with the cells bordering this
   region.}
   \label{fig:regions_pcisph}
 \end{figure}

\subsection{Global Density Correction Schedule}
\begin{figure}
   \centering
   \includegraphics[width=1.0\linewidth]{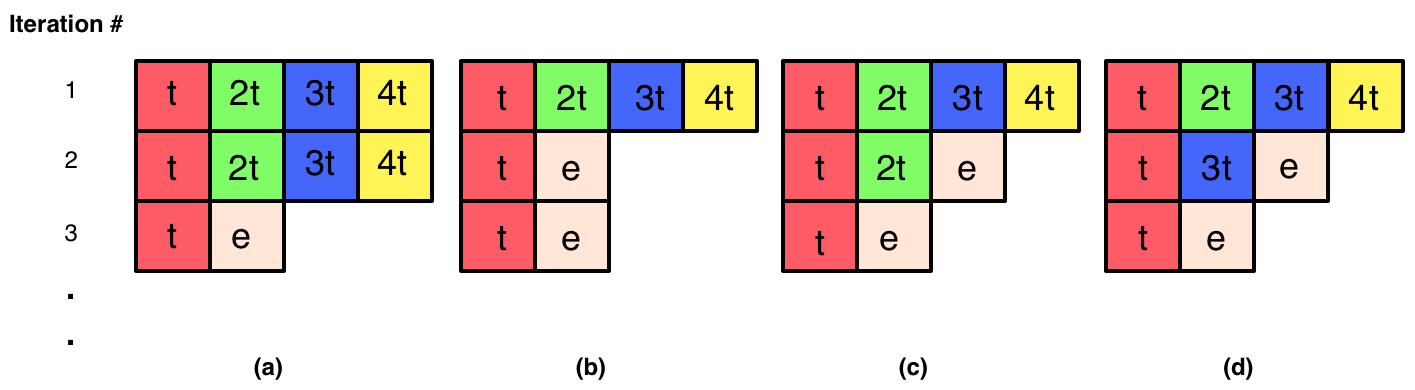}
   \caption{Our density correction schedule for PCISPH.  Each column (a)-(d) represents one minor time step within a major time step of $4t$. Each row within a column represents one iteration
   of density correction being applied. The colored blocks within each row indicate which regions have density correction performed on them at a given iteration within a minor step. Blocks
   labelled $\mathbf{e}$ correspond to iterations performed for all remaining particles whose density error hasn't been fully corrected. \newline The schedule we designed has the following
   properties. All regions receive a minimum of one iteration of density correction on every step, as seen in the first row, with the exception of the very first step within a major step,
   in which all regions receive two iterations. Additional iterations are then assigned based on region membership. The fastest moving ($t$) regions receive 3 iterations per step, as
   in classic PCISPH. The next fastest moving ($2t$) region receives 3 iterations for every two minor steps. Slower moving regions similarly receive fewer total iterations per major step. }
   \label{fig:iterations_pcisph}
    \end{figure}

On each minor time step, we perform a certain number of iterations of density correction which varies by region. The number of iterations applied to each region type is determined by
the density correction schedule shown in Figure \ref{fig:iterations_pcisph}. This schedule which was chosen heuristically to satisfy certain constraints. Specifically, a minimum of one
correction iteration should be applied to all particles to keep them minimally synchronized. Within the ``true" time step for a given particle it should cumulatively receive at least 3
correction iterations (similar to standard PCISPH). Over a given major step, slower and faster moving regions should undergo fewer and more iterations, respectively. The first minor step
within a major step begins with all regions undergoing two iterations rather than just one. This schedule was effective in all the scenarios we tried, though other choices may be possible.
Within our density correction algorithm (Algorithm \ref{alg:DENSITYCORR_RTS}) the variable $turn$ is set to true if a given particle is scheduled to undergo density correction on the current
iteration.

\subsection{Local Density Correction}
 We sometimes find that a small number of particles have unresolved density errors after their assigned number of correction iterations is completed on a given step. We address this in a
 manner similar to Raveendran et al.\cite{Raveendran:2011:HSP:2019406.2019411}.

We introduce an additional region  $\Re_e$ (with particle set $S_e$); this constitutes all particles having error larger than a chosen threshold irrespective of their time step.
We follow the same error metric as given by Ihmsen et al.\cite{IhmsenAGT10} where the average density error $\rho^{avg}_{err}$ of all particles and the maximum density error
$\rho^{max}_{err}$ of any particle should not exceed a certain value. Per Raveendran et al., this threshold is set to half the value of the largest allowed density error. The particles
that compose $S_e$ are corrected at every iteration of our schedule. If no such particles exist, we skip this step.

Since we often perform density correction on just a subset of all the particles, we need to know if these corrections have disrupted the density of particles whose density was previously
declared correct. Movement near or across the boundaries between regions can easily cause this to occur. To account for this we identify another region, $\Re_{ob}$, which is the set of
"observed" blocks that we monitor for changes, as shown in Figure \ref{fig:regions_pcisph}. This is the set of blocks that border on either $\Re_{e}$ or a region with a smaller time step;
its particle set is denoted $S_{ob}$. After each iteration, it suffices to check just the density of particles in $S_{ob}$ for newly introduced errors, avoiding an expensive global check
on all particles whose density is already correct.

\subsection{Extra Correction Iterations}
After the standard 3 iteration schedule is completed for a given minor step, if some error in density remains we do one of two things. We can either perform additional purely
\emph{local} corrections on the particles with remaining error, or run through additional \emph{global} iterations by resetting to the top of the correction schedule for the current minor
step. Local correction is often preferable since it minimizes the number of particles involved, but it may not always be efficient or feasible if the error is large or global. We therefore
make this choice by considering whether the average compression error $\rho^{avg}_{err}$ exceeds a threshold, since this is indicative of global density errors. If local iterations are
initially selected, but they nonetheless fail to converge after a few iterations, we switch back to global corrections
and revert the pressure estimates to their values from before local iterations began.

If a minor step exceeds 6 global iterations for correction, we terminate the major step early and temporarily reduce the length of subsequent major steps to $2\Delta t_b$ (i.e., two minor
steps rather than four). This allows the global neighborhood grid to be completely updated more frequently to better handle these large density changes. After 10 major steps in which we do
not exceed 6 global iterations on any minor steps, the major time step is reverted.  However, we emphasize that in our experiments we found that for the majority of cases extra global
operations aren't necessary; a few steps of local correction typically suffice.

For PCISPH we make one additional minor change to our region determination strategy: we expand regions $\Re_1$ and $\Re_2$ by $4$ layers of blocks each before beginning a major step,
in a manner similar to $Re_{min}$. This is done to account for fast particles that may travel several blocks from their original position.

\subsection{Neighborhood Computation and Updates}

At the start of each major step, we compute the block grid and the particles contained in each block. Particles then compute their neighborhood using this grid, along with their region
membership and the number of minor steps for which they are considered valid. This neighborhood will be used to compute external forces and the pressures needed to correct density variations. Since block updates and neighborhood search are generally expensive operations, we prefer to avoid them as much as possible, through the use of an asynchronous neighborhood update strategy.
Particles assigned to faster regions update their neighborhoods whenever their validity expires, at the start of a minor step. That is, particles in region $\Re_1$ update their neighborhood
every step, particles in $\Re_2$ update every second minor step, and particles in $\Re_4$ do so only once per major time step. We found that since particles in $\Re_3$ are also slow moving,
postponing their neighborhood update to the next major step is satisfactory as well.

At each minor step, we also check if a particle now requires a smaller time step. If so, we mark its block and the neighboring blocks to update their time step. This allows the simulation
to adapt to sudden changes.

   \begin{figure}[t]
   \centering
   \includegraphics[width=.8\linewidth]{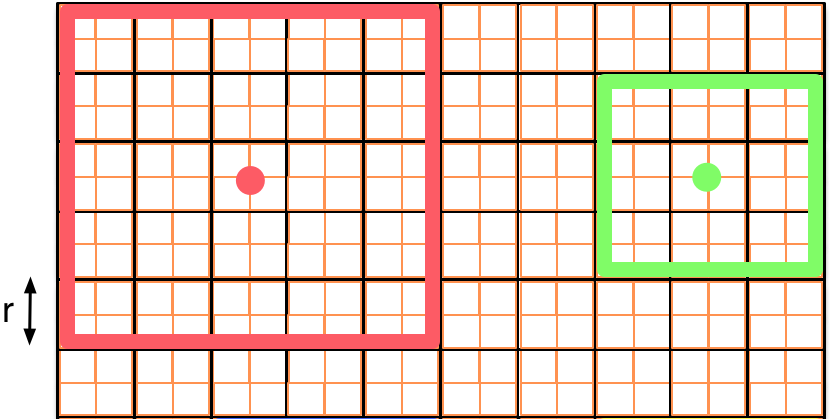}
   \caption{For PCISPH particles in $\Re_1$, we search two layers of grid cells (red) when determining its particle neighborhood, rather than the usual one (green).}
   \label{fig:neigh_pcisph}
    \end{figure}

When a particle's validity expires and its neighborhood needs updating, a na\"{i}ve approach would globally update the particles belonging to each grid block and performing a new search
within the updated neighboring blocks. This would be very costly, limiting the benefit of our asynchronous strategy. To cope with this, we take two steps. First, we initially overpopulate
the grid at the start of each major step; that is, we use a grid with blocks of size $r = 2.2s$ rather than $r = 2s$. This ensures that particles will nearly always find all the particles
belonging to their neighborhood in their own or adjacent blocks, even if we keep the same block data over the entire major step.

\begin{figure}[t]
   \centering
    \subfigure{ \includegraphics[width=.35\linewidth]{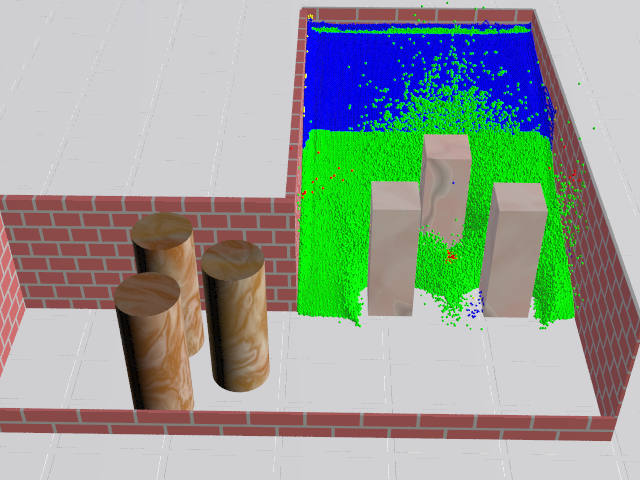} \label{}
                \includegraphics[width=.35\linewidth]{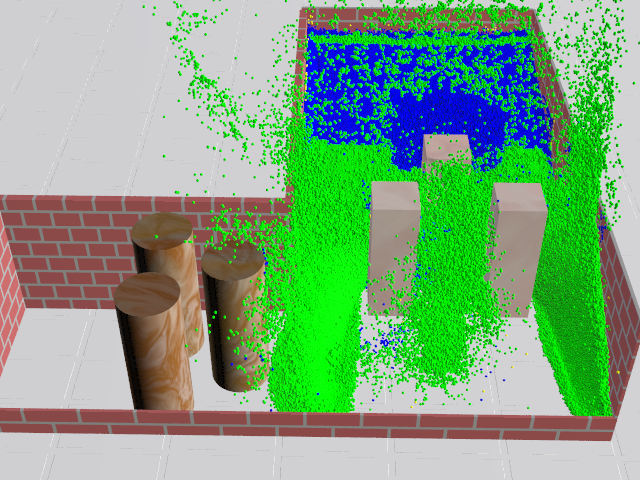} \label{}
                \includegraphics[width=.35\linewidth]{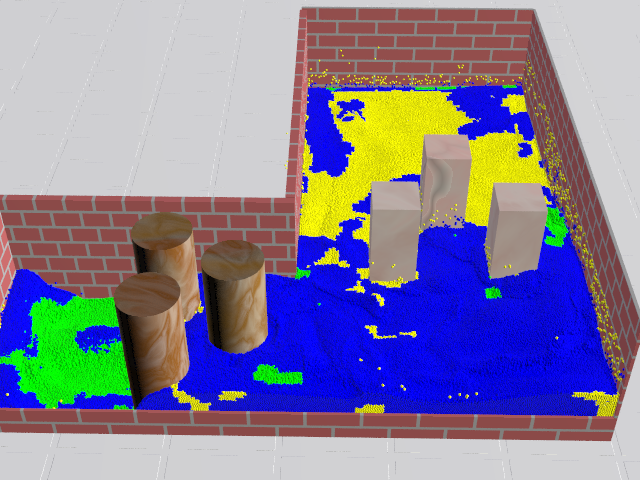} \label{}}
   \caption{Extending the neighborhood search radius to two blocks for particles in both $\Re_1$ and $\Re_2$ results in little evident difference when compared to \ref{fig:gallery_rts},
   in which only the search radius for $\Re_1$ is expanded.}
   \label{fig:pcisph_ext_comparison}
\end{figure}

Second, for the very fast particles within $\Re_1$, we expand the radius of our neighbor particle search to two grid blocks instead of one (see Figure \ref{fig:neigh_pcisph}). While this
means we must check many more possible neighbor particles, this is justified because these particles are few in number and arise infrequently, yet are critical to stability. Doing this with
the particles in $\Re_2$ as well would be substantially more expensive. Figure \ref{fig:pcisph_ext_comparison} illustrates that expanding the search radius for both $\Re_1$ and $\Re_2$ makes
little visual difference, so we avoid doing so for the sake of efficiency. (Moreover, standard PCISPH assumes a fixed neighborhood over large time steps
\cite{Solenthaler:2009:PIS:1531326.1531346}, so the method is reasonably robust to small errors in neighborhood estimates.)

\begin{algorithm}
\begin{algorithmic}[1]
\WHILE{animating}
    \STATE{update block neighborhood grid} \newline

     \STATE{\textcolor{blue}{/*---------- Region determination ($\Re_i$)  ---------*/}}

     \FORALL{$i$ $\in$ $S$}
        \STATE {update parent block maximum with $v_i$ and $F_i^{total}$} \newline
     \ENDFOR

    \FORALL{blocks $b$}
        \STATE {compute new region membership } \newline
     \ENDFOR
     \STATE{expand regions $\Re_1$, $\Re_2$}

     \STATE{find $observed$ blocks $R_{ob}$ and particle set $S_{ob}$}

    \FORALL{$i$ $\in$ $S$}
        \STATE{update $timestep_i$, $validity_i$, $compute_i$ using parent block} \newline
    \ENDFOR

     \STATE{\textcolor{blue}{/*---------- $N$ Minor time steps ---------*/}}
     \FOR{j = 1 to N}

      \FORALL{$i$ $\in$ $S$}
            \LINEIF{$compute_i$}{update neighborhood $N_{i}$} \newline
        \ENDFOR

        \FORALL{$i$ $\in$ $S$}
            \LINEIF{$compute_i$}{update $\textbf{F}^{v, g, ext}$} \newline
        \ENDFOR

        \FORALL{$i$ $\in$ $S$}
            \STATE{initialize pressure $p(t)$ = 0.0}
            \STATE{initialize pressure force $\textbf{F}^{p(t)}$ = 0.0}  \newline
        \ENDFOR

        \STATE{\textcolor{blue}{/*---------- Density Correction ---------*/}}
        \STATE{DensityCorrectionRTS}    \newline

        \FORALL{$i$ $\in$ $S$}
            \STATE {compute new $v_i(t+1)$}
            \STATE {compute new $x_i(t+1)$} \newline
        \ENDFOR


        \FORALL{$i$ $\in$ $S$}
            \STATE{decrement $validity_i$}
            \IF {($validity_i$ $\leq$ $0$)}
                \STATE {set $compute_i$ $=$ \textbf{true}}
                \STATE {update $validity_i$}
             \ELSE
                \STATE {$compute_i$ $=$ \textbf{false}} \newline
             \ENDIF
        \ENDFOR

     \ENDFOR
\ENDWHILE

\end{algorithmic}
\caption{Regional Time Stepping for PCISPH}
\label{alg:PCISPH_RTS}
\end{algorithm}

\begin{algorithm}
\begin{algorithmic}[1]

\STATE{set $minIterations$ to 3}
\STATE{set $S_e$ to \textbf{NULL}}  \newline
\WHILE{($iter$ $\le$ $minIterations$) $\|$ ($\rho^{*}_{err}(t+1)$ $\ge$ $\eta$)}
      \FORALL{$i$ $\in$ $S$}
            \STATE{predict velocity $v_i^*(t+1)$}
            \STATE{predict position $x_i^*(t+1)$}  \newline
      \ENDFOR

      \FORALL{$i$ $\in$ $S$}
            \IF{($i$ has turn) $\|$ ($i$ $\in$ $S_e$) $\|$ ($i$ $\in$ $S_{ob}$)}
                \STATE{predict density $\rho_i^*(t+1)$}
                \STATE{update density variation $\rho^*_{err}(t+1)$}
                \STATE{update pressure $p_i(t)$ += $f(\rho^*_{err}(t+1))$} \newline
            \ENDIF
      \ENDFOR

      \IF{(\textit{density error remains outside active regions})}
        \STATE{update $S_e$, $S_{ob}$}  \newline
      \ENDIF

       \FORALL{$i$ $\in$ $S$}
            \IF{($i$ has turn) $\|$ ($i$ $\in$ $S_e$) }
                \STATE{compute pressure force $\textbf{F}_i^p(t+1)$} \newline
            \ENDIF
      \ENDFOR

      \STATE{$iter$++}

     \IF{(\textit{$\rho^{*}_{err}(t+1)$ $\ge$ $\eta$}) $\&$ ($iter$ $\ge$ 3 )}
        \STATE{$minIterations$++}
        \IF{($\rho^{avg}_{err}$ $\ge$  $\eta_{T}$)}
            \STATE{reset to top of active step's correction schedule (i.e., extra global corrections)}
        \ELSE
            \STATE{perform local correction on $S_e$}
        \ENDIF
      \ENDIF

\STATE{}
\ENDWHILE{}

\end{algorithmic}
\caption{DensityCorrectionRTS}
\label{alg:DENSITYCORR_RTS}
\end{algorithm}

\section{Results}

The proposed and baseline WCSPH/PCISPH methods were implemented on a MAC 10.7.5 machine with 3.2 GHz quad-core Intel processor, using C++ and the OpenMP API.
The images were rendered offline with POVRAY.

For WCSPH, we have used $\Delta t_{cfl}$ for standard time stepping and $\Delta t_b = \Delta t_{cfl}$ for RTS.
For PCISPH, we compare global time stepping with a constant time step against both adaptive global stepping \cite{IhmsenAGT10} and our method.
For the sake of experimental similarity with Ihmsen's results, we have used a constant time step of $0.000166$ for standard PCISPH method and no static boundary particles. For both
global time stepping PCISPH and RTS PCISPH we use static boundary particles. Inactive static boundary particles are culled prior to computing the physics.

\begin{figure}[t]
   \centering
    \subfigure{ \includegraphics[width=.32\linewidth]{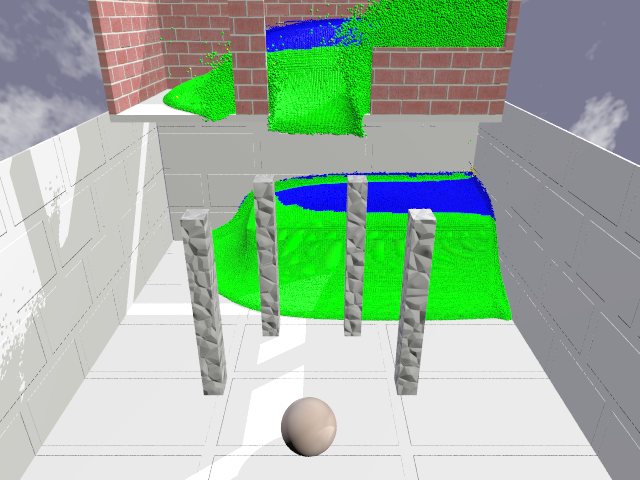} \label{}
                \includegraphics[width=.32\linewidth]{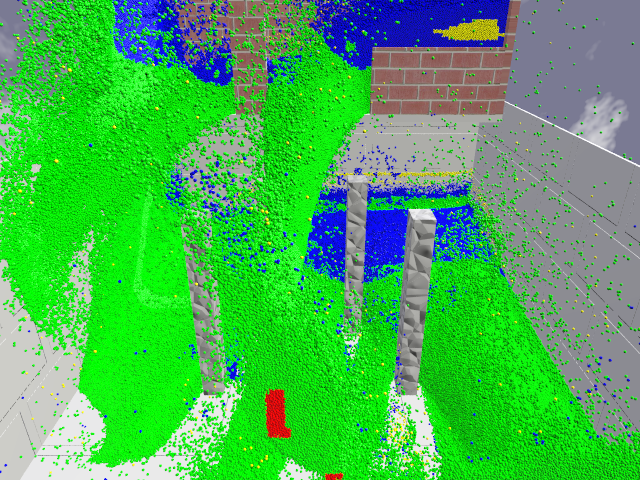} \label{}
                \includegraphics[width=.32\linewidth]{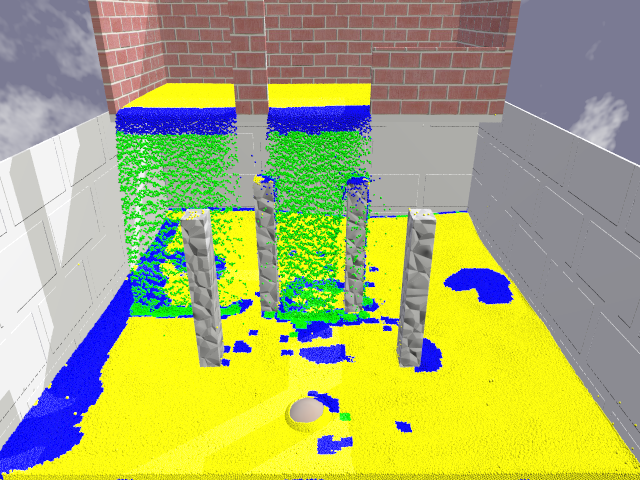}  \label{}}
  \subfigure{ \includegraphics[width=.32\linewidth]{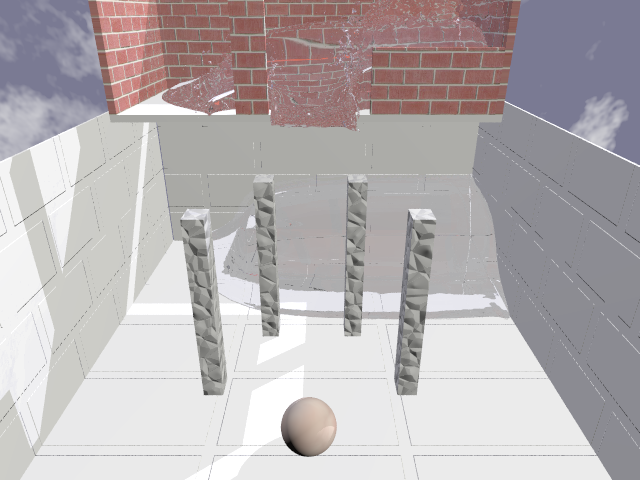} \label{}
                \includegraphics[width=.32\linewidth]{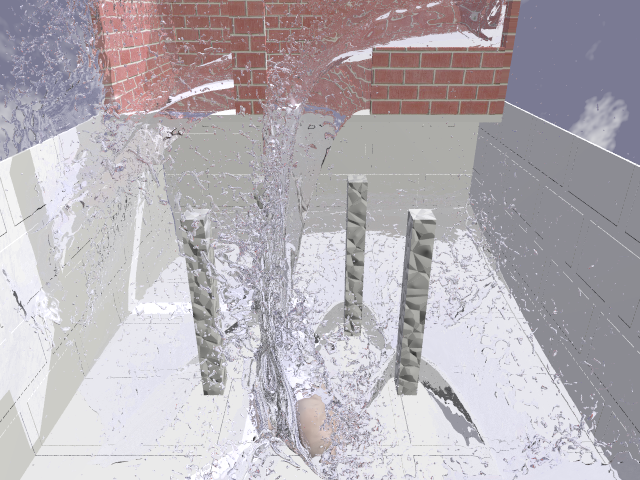} \label{}
                \includegraphics[width=.32\linewidth]{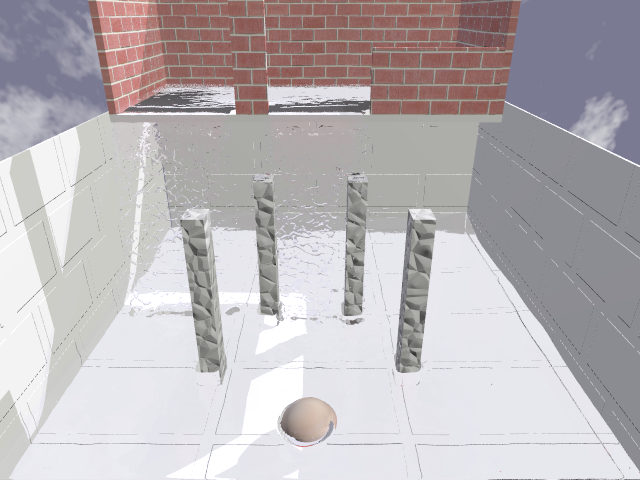}  \label{}}
  \caption{A highly turbulent flow involving several obstacles and boundaries, simulated with 4.5M particles using the proposed regional time stepping algorithm for PCISPH. }
   \label{fig:pcisph_turbine}
\end{figure}

\begin{figure}[t]
   \centering
    \subfigure[]{ \includegraphics[width=.35\linewidth]{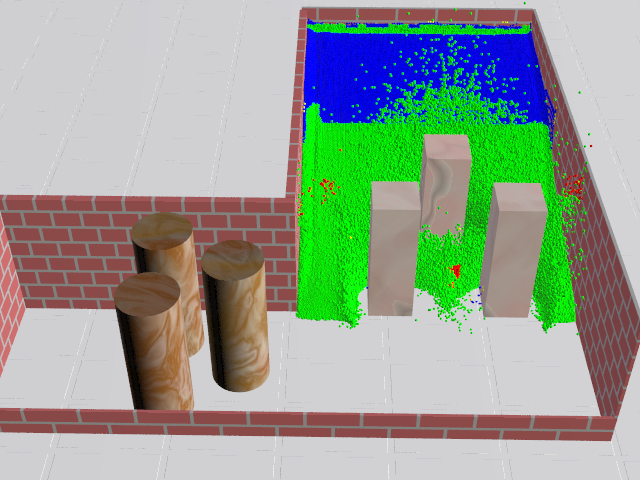} \label{}
                \includegraphics[width=.35\linewidth]{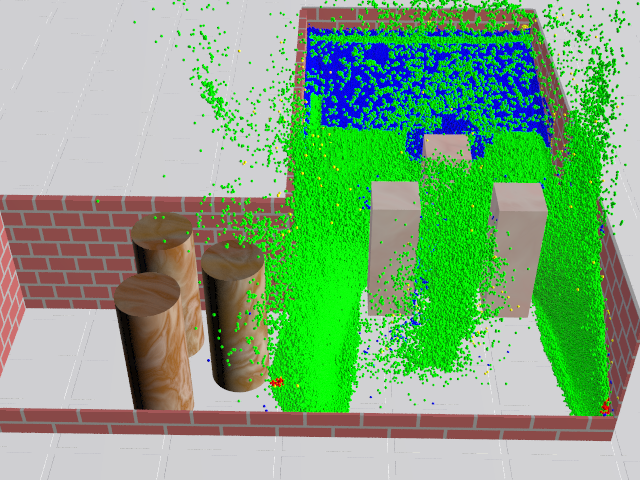} \label{}
                \includegraphics[width=.35\linewidth]{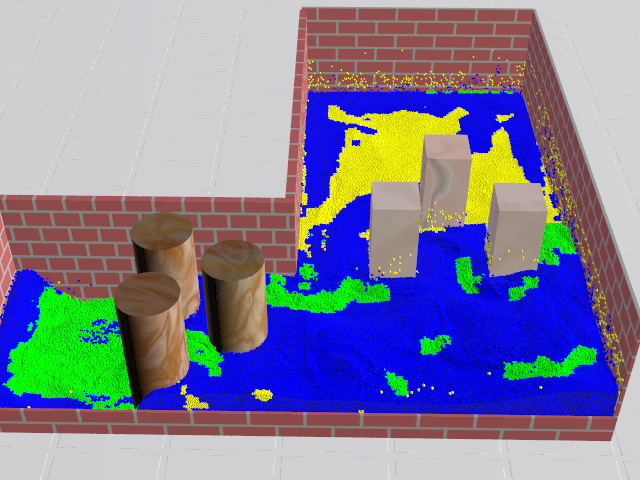}  \label{fig:gallery_rts}}
    \subfigure[]{ \includegraphics[width=.35\linewidth]{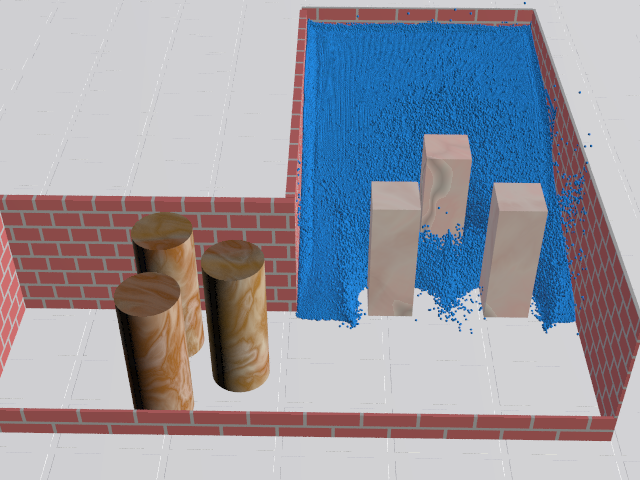} \label{}
                \includegraphics[width=.35\linewidth]{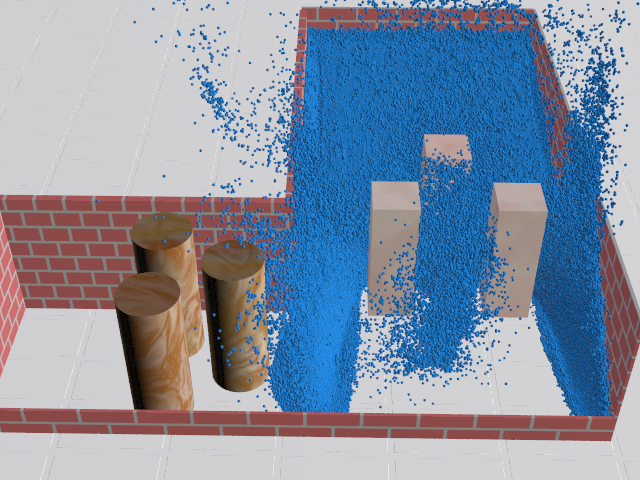} \label{}
                \includegraphics[width=.35\linewidth]{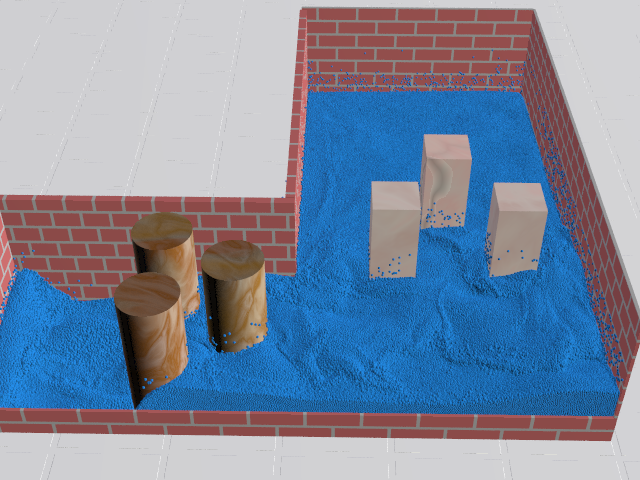} \label{}}
    \subfigure[]{ \includegraphics[width=.35\linewidth]{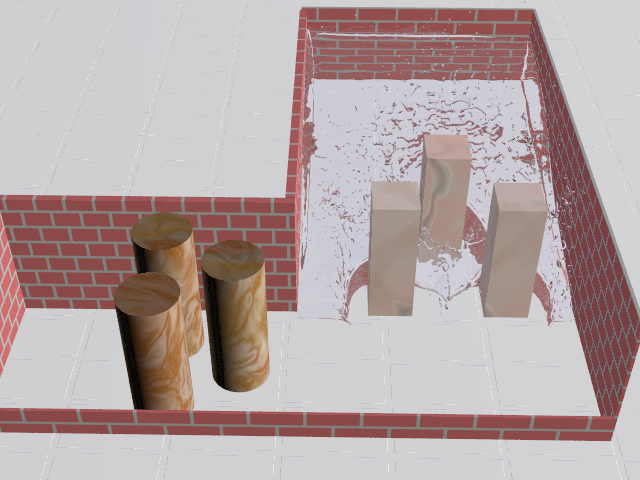} \label{}
                \includegraphics[width=.35\linewidth]{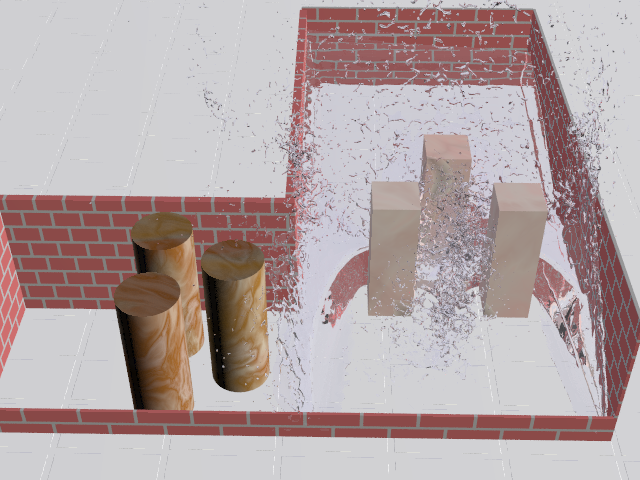} \label{}
                \includegraphics[width=.35\linewidth]{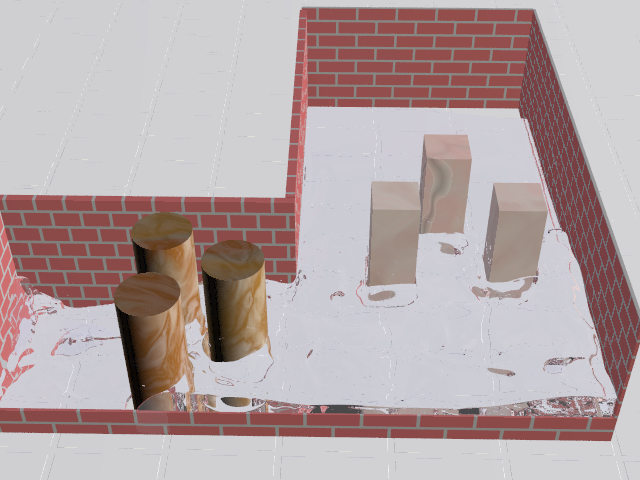} \label{}}
    \subfigure[]{ \includegraphics[width=.35\linewidth]{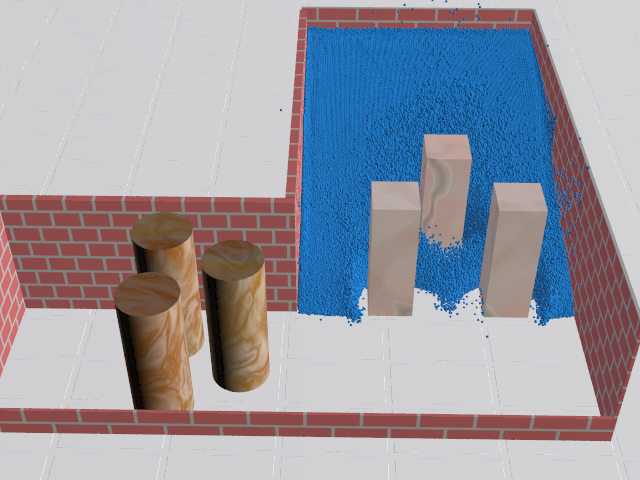} \label{}
                \includegraphics[width=.35\linewidth]{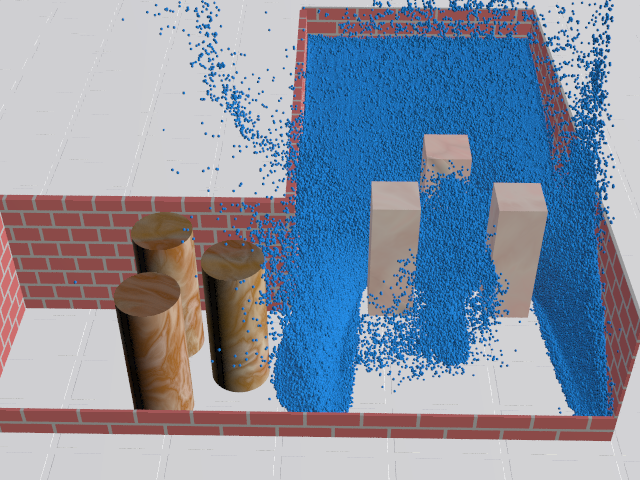} \label{}
                \includegraphics[width=.35\linewidth]{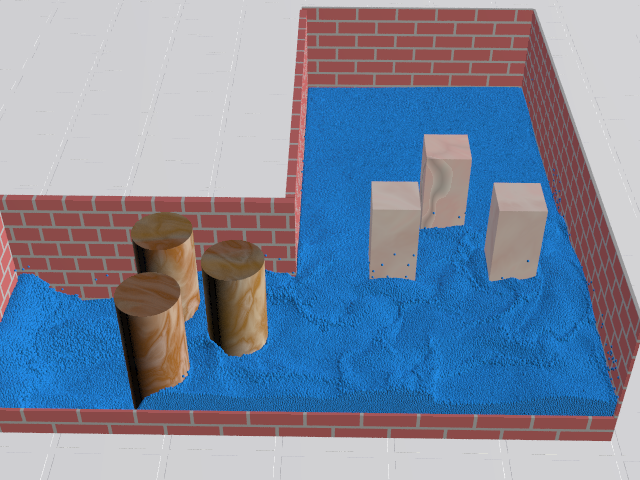} \label{}}
    \subfigure[]{ \includegraphics[width=.35\linewidth]{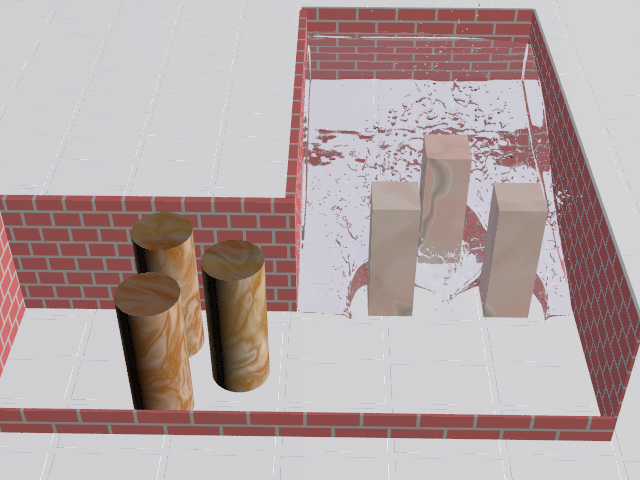} \label{}
                \includegraphics[width=.35\linewidth]{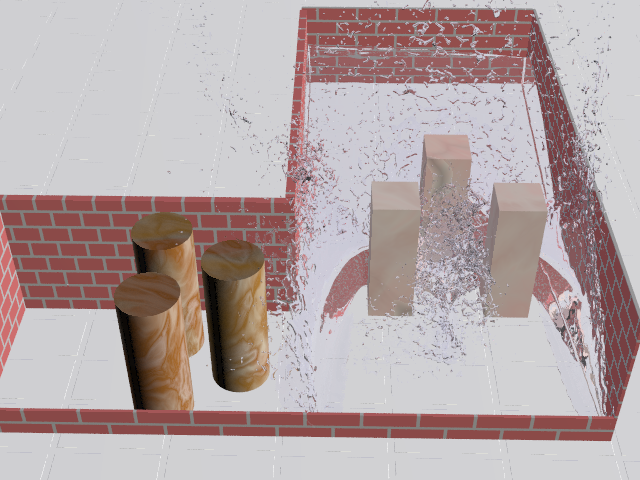} \label{}
                \includegraphics[width=.35\linewidth]{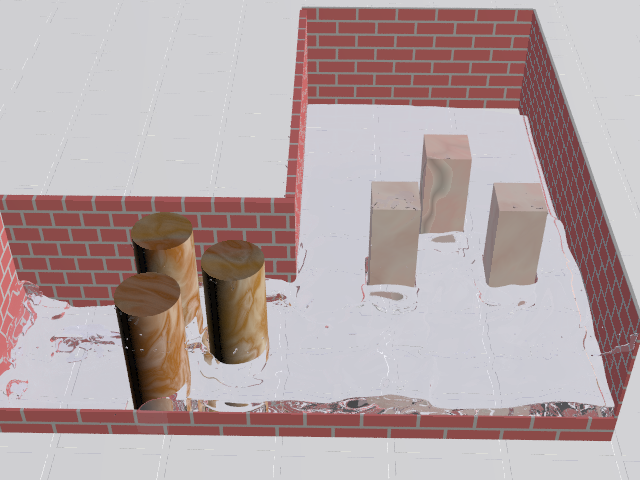} \label{}}

   \caption{RTS PCISPH compared with standard PCISPH. Our method produces overall behavior and features close to the original. (a) RTS particles colored according to their time steps,
   (b) RTS particles uniformly colored for comparison purpose, (c) RTS surface for corresponding frames, (d) standard particles and (e) standard surface for corresponding frames. }
   \label{fig:pcisph_comparison}
\end{figure}

For the scenes in Table \ref{table:performTable}, $\Delta t_b$ is chosen to be $0.00035$ for Figure \ref{fig:pcisph_corridor} and \ref{fig:pcisph_comparison}, and $0.0003$ for Figure
\ref{fig:pcisph_turbine}. This essentially allows a larger time step of $4 \Delta t_b$ in each major cycle. The user-chosen error threshold $\rho_T$ (for whether to perform extra global
or local corrections) is set to $0.25$ in all our experiments.

Since we a moderately restrictive value of $\alpha=0.4$ in Equation \ref{eq:time} for RTS WCSPH, the region variation is more sensitive, particularly between regions $\Re_2$ and $\Re_3$.
For the same reason, particles do not often take time steps larger than $3\Delta t_b$. Our block-based RTS algorithm allows these transitions regions and occasional very fast moving
particles to be treated efficiently, while keeping the simulation stable and accurate.

Table \ref{table:performTable} gives the performance speed-up of our methods in comparison to both WCSPH and PCISPH for several examples. Our methods yield simulations between 1.7 and 2.1
times faster than the comparison methods for the given examples. Table \ref{table:stats} provides relative timing breakdowns for our methods into physics computations, neighborhood searching,
 and regional time-stepping components. Figure \ref{fig:pcisph_comparison} and our video examples illustrate that RTS approaches yield results that are visually consistent with the original
  SPH schemes. Furthermore, they work well even for challenging dynamic scenes involving frequent collisions with boundaries and obstacles.

\begin{figure}[t]
   \centering
    \subfigure{ \includegraphics[width=.3\linewidth]{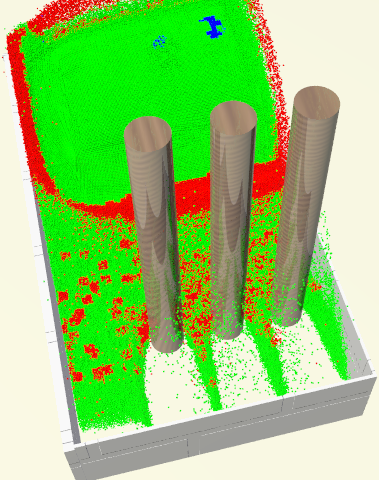} \label{}
                \includegraphics[width=.3\linewidth]{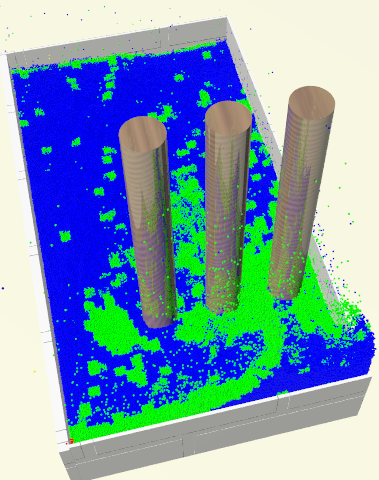} \label{}
                \includegraphics[width=.3\linewidth]{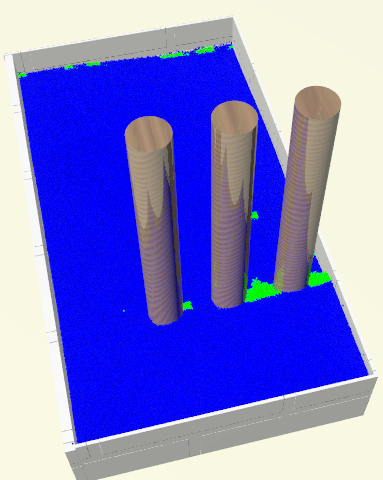}  \label{}}
    \subfigure{ \includegraphics[width=.3\linewidth]{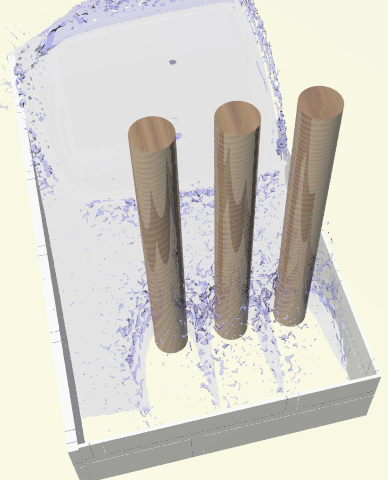} \label{}
                \includegraphics[width=.3\linewidth]{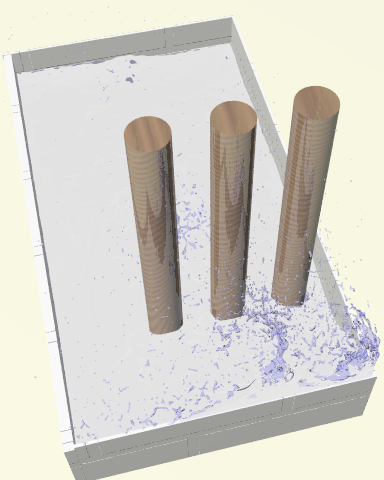} \label{}
                \includegraphics[width=.3\linewidth]{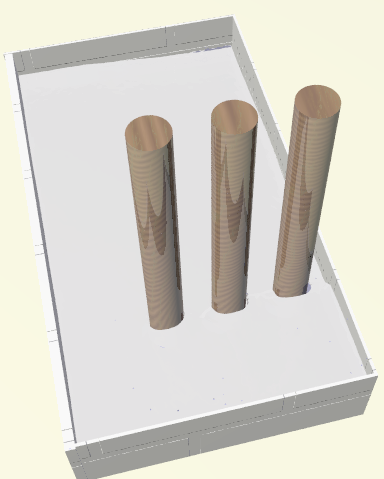}  \label{}}
  \caption{A dam break scenario where immediately after releasing the dam we drop a second block of liquid on top, simulated with 2.2M particles using RTS WCSPH.}
   \label{fig:sph_blockDrop}
\end{figure}

\begin{table}[ht]
\centering  
\small
\begin{tabular}{|l|c|c|c|c|}
\hline
\cline{1-5}
 & Scene &  \# particles & Speed up& Speed up\\
 \hline \hline
        &        & & - & Ours vs. Standard \\
 \cline{4-5}
        & Figure \ref{fig:sph_doubleDB} & 1.5M & - & 2\\
  SPH   &        & & &\\
        & Figure \ref{fig:sph_blockDrop} & 2.2M & - & 1.8\\
        &        & & & \\

 \hline
       &        & & Ihmsen vs. Constant & Ours vs. Constant \\
 \cline{4-5}
            & Figure \ref{fig:pcisph_corridor}& 1.6M &  2.1 & 3.6 \\
            &        & & &\\
 PCISPH     & Figure \ref{fig:pcisph_comparison}& 2.3M & 2.3 & 4.9 \\
            &        & & &\\
            & Figure \ref{fig:pcisph_turbine}& 4.5M & 2.2 & 4.3 \\
            &        & & &\\
 \hline
\end{tabular}
\caption{Performance comparisons of regional time stepping with other methods.} 
\label{table:performTable}
\end{table}

\begin{table}[ht]
\centering
\begin{tabular}{|r|r|r|r|}
\hline
$ $ & $T_{neighbor}$ & $T_{physics}$ & $T_{RTS}$\\
\hline \hline
WCSPH & 37\% & 42\% & 14\% \\ 
PCISPH & 28\% & 56\% & 12\% \\
\hline
\end{tabular}
\caption{Timing breakdowns for each method. Computations specific to RTS constitute only 10-15\% of the run-time.}
\label{table:stats}
\end{table}

\section{Conclusions and Future Work}

We have presented an efficient technique for regional time-stepping for WCSPH and PCISPH. The proposed methods are simple to implement and can achieve significant speed-ups while maintaining
behaviour consistent with the corresponding fully synchronous simulations. We envision several directions for future work. Two natural extensions would be implementing a GPU-based version,
and combining our method with spatial adaptivity, such as Solenthaler's two-scale method \cite{Solenthaler:2011:TPS:2010324.1964976}. Another promising research direction could be to integrate
regional time stepping with implicit incompressible SPH \cite{Ihmsen:2014:IIS:2574216.2574356}, position based fluids \cite{Macklin:2013:PBF:2461912.2461984} and divergence free SPH \cite{Bender2015}
It could also be useful to adjust the base step of our PCISPH scheme in a manner similar to the adaptive global time stepping method of Ihmsen et al.\cite{IhmsenAGT10}, in order to
achieve the benefits of both. Finally, to reap the full benefits of asynchrony in a wider range of scenarios we would like to explore methods that can couple between SPH and other
asynchronously integrated physical systems, such as deformable or rigid bodies.

\section{Conflict of interest}
All persons who meet authorship criteria are listed as authors, and all authors certify that they have
participated sufficiently in the work to take public responsibility for the content, including participation
in the concept, design, analysis, writing, or revision of the manuscript. Furthermore, each author
certifies that this paper (and its contents) is original research work and has not been and will not be 
submitted to or published in any other publication.


\end{document}